\documentclass[%
 reprint,
superscriptaddress,
showpacs,
nofootinbib,
nobibnotes,
amsmath,amssymb,
aps,
prl, 
longbibliography,
twocolumn
]{revtex4-2}

\usepackage{graphicx}
\usepackage{dcolumn}
\usepackage{bm}
\usepackage{xcolor}
\usepackage[caption=false]{subfig}
\usepackage{lineno}
\usepackage{braket}

\begin{document}


\title{First Search for Dark Sector $\mathbf{e^+e^-}$ Explanations of the MiniBooNE Anomaly at MicroBooNE}


\newcommand{\bologna}{ Dipartimento di Fisica e Astronomia, Universita` di Bologna, Bologna, Italy }
\newcommand{\CERN}{Information Technology Department, CERN, 1211 Geneva 23, Switzerland}
\newcommand{\Harvard}{Harvard University, Cambridge, MA 02138, USA}
\newcommand{\INFN}{ INFN, Sezione di Bologna, Bologna, Italy}
\newcommand{\LOUVAIN}{Centre for Cosmology, Particle Physics and Phenomenology - CP3, Université catholique de Louvain, Louvain-la-Neuve, Belgium}
\newcommand{\ANL}{Argonne National Laboratory (ANL), Lemont, IL, 60439, USA}
\newcommand{\Bern}{Universit{\"a}t Bern, Bern CH-3012, Switzerland}
\newcommand{\BNL}{Brookhaven National Laboratory (BNL), Upton, NY, 11973, USA}
\newcommand{\UCSB}{University of California, Santa Barbara, CA, 93106, USA}
\newcommand{\Cambridge}{University of Cambridge, Cambridge CB3 0HE, United Kingdom}
\newcommand{\CIEMAT}{Centro de Investigaciones Energ\'{e}ticas, Medioambientales y Tecnol\'{o}gicas (CIEMAT), Madrid E-28040, Spain}
\newcommand{\Chicago}{University of Chicago, Chicago, IL, 60637, USA}
\newcommand{\Cincinnati}{University of Cincinnati, Cincinnati, OH, 45221, USA}
\newcommand{\CSU}{Colorado State University, Fort Collins, CO, 80523, USA}
\newcommand{\Columbia}{Columbia University, New York, NY, 10027, USA}
\newcommand{\Edinburgh}{University of Edinburgh, Edinburgh EH9 3FD, United Kingdom}
\newcommand{\FNAL}{Fermi National Accelerator Laboratory (FNAL), Batavia, IL 60510, USA}
\newcommand{\Granada}{Universidad de Granada, Granada E-18071, Spain}
\newcommand{\IIT}{Illinois Institute of Technology (IIT), Chicago, IL 60616, USA}
\newcommand{\IFT}{Instituto de Fisica Teorica (IFT), Madrid  E-28049, Spain }
\newcommand{\ICL}{Imperial College London, London SW7 2AZ, United Kingdom}
\newcommand{\Indiana}{Indiana University, Bloomington, IN 47405, USA}
\newcommand{\Kansas}{The University of Kansas, Lawrence, KS, 66045, USA}
\newcommand{\KSU}{Kansas State University (KSU), Manhattan, KS, 66506, USA}
\newcommand{\Lancaster}{Lancaster University, Lancaster LA1 4YW, United Kingdom}
\newcommand{\LANL}{Los Alamos National Laboratory (LANL), Los Alamos, NM, 87545, USA}
\newcommand{\Louisiana}{Louisiana State University, Baton Rouge, LA, 70803, USA}
\newcommand{\Manchester}{The University of Manchester, Manchester M13 9PL, United Kingdom}
\newcommand{\MIT}{Massachusetts Institute of Technology (MIT), Cambridge, MA, 02139, USA}
\newcommand{\Michigan}{University of Michigan, Ann Arbor, MI, 48109, USA}
\newcommand{\MSU}{Michigan State University, East Lansing, MI 48824, USA}
\newcommand{\Minnesota}{University of Minnesota, Minneapolis, MN, 55455, USA}
\newcommand{\Nankai}{Nankai University, Nankai District, Tianjin 300071, China}
\newcommand{\NCNR}{National Centre for Nuclear Research, ul. Pasteura 7, 02-093 Warsaw, Poland}
\newcommand{\NMSU}{New Mexico State University (NMSU), Las Cruces, NM, 88003, USA}
\newcommand{\Oxford}{University of Oxford, Oxford OX1 3RH, United Kingdom}
\newcommand{\Pitt}{University of Pittsburgh, Pittsburgh, PA, 15260, USA}
\newcommand{\QMUL}{Queen Mary University of London, London E1 4NS, United Kingdom}
\newcommand{\Rutgers}{Rutgers University, Piscataway, NJ, 08854, USA}
\newcommand{\SLAC}{SLAC National Accelerator Laboratory, Menlo Park, CA, 94025, USA}
\newcommand{\SDSMT}{South Dakota School of Mines and Technology (SDSMT), Rapid City, SD, 57701, USA}
\newcommand{\Maine}{University of Southern Maine, Portland, ME, 04104, USA}
\newcommand{\Syracuse}{Syracuse University, Syracuse, NY, 13244, USA}
\newcommand{\TelAviv}{Tel Aviv University, Tel Aviv, Israel, 69978}
\newcommand{\UTA}{University of Texas, Arlington, TX, 76019, USA}
\newcommand{\Tufts}{Tufts University, Medford, MA, 02155, USA}
\newcommand{\VTech}{Center for Neutrino Physics, Virginia Tech, Blacksburg, VA, 24061, USA}
\newcommand{\Warwick}{University of Warwick, Coventry CV4 7AL, United Kingdom}

\affiliation{\ANL}
\affiliation{\Bern}
\affiliation{\BNL}
\affiliation{\bologna}
\affiliation{\UCSB}
\affiliation{\Cambridge}
\affiliation{\CERN}
\affiliation{\CIEMAT}
\affiliation{\Chicago}
\affiliation{\Cincinnati}
\affiliation{\CSU}
\affiliation{\Columbia}
\affiliation{\Edinburgh}
\affiliation{\FNAL}
\affiliation{\Granada}
\affiliation{\Harvard}
\affiliation{\IIT}
\affiliation{\ICL}
\affiliation{\IFT}
\affiliation{\Indiana}
\affiliation{\INFN}
\affiliation{\Kansas}
\affiliation{\KSU}
\affiliation{\Lancaster}
\affiliation{\LANL}
\affiliation{\Louisiana}
\affiliation{\LOUVAIN}
\affiliation{\Manchester}
\affiliation{\MIT}
\affiliation{\Michigan}
\affiliation{\MSU}
\affiliation{\Minnesota}
\affiliation{\Nankai}
\affiliation{\NCNR}
\affiliation{\NMSU}
\affiliation{\Oxford}
\affiliation{\Pitt}
\affiliation{\QMUL}
\affiliation{\Rutgers}
\affiliation{\SLAC}
\affiliation{\SDSMT}
\affiliation{\Maine}
\affiliation{\Syracuse}
\affiliation{\TelAviv}
\affiliation{\UTA}
\affiliation{\Tufts}
\affiliation{\VTech}
\affiliation{\Warwick}

\author{A.~M.~Abdullahi} \affiliation{\IFT}   
\author{P.~Abratenko} \affiliation{\Tufts}
\author{D.~Andrade~Aldana} \affiliation{\IIT}
\author{L.~Arellano} \affiliation{\Manchester}
\author{J.~Asaadi} \affiliation{\UTA}
\author{A.~Ashkenazi}\affiliation{\TelAviv}
\author{S.~Balasubramanian}\affiliation{\FNAL}
\author{B.~Baller} \affiliation{\FNAL}
\author{A.~Barnard} \affiliation{\Oxford}
\author{G.~Barr} \affiliation{\Oxford}
% ...

\date{\today}

\begin{abstract}
We present MicroBooNE's first search for dark sector $e^+e^-$ explanations of the long-standing MiniBooNE anomaly. The MiniBooNE anomaly has garnered significant attention over the past 20 years including previous MicroBooNE investigations into both anomalous electron and photon excesses, but its origin still remains unclear. In this letter, we provide the first direct test of dark sector models in which dark neutrinos, produced through neutrino-induced scattering, decay into missing energy and visible $e^+e^-$ pairs comprising the MiniBooNE anomaly. Many such models have recently gained traction as a viable solution to the anomaly while evading past bounds. Using an exposure of $6.87 \times 10^{20}$ protons-on-target in the Booster Neutrino Beam, we implement a selection targeting forward-going, coherently produced $e^+e^-$ events. After unblinding, we observe 95 events, which we compare with the constrained background-only prediction of $69.7 \pm 17.3$. This analysis sets the world’s first direct limits on these dark sector models and, at the 95\% confidence level, excludes the majority of the parameter space viable as a solution to the MiniBooNE anomaly.
\end{abstract}

\maketitle

The anomalous results obtained in the short-baseline MiniBooNE~\cite{MiniBooNE:2008yuf,MiniBooNE:2018esg,Aguilar_Arevalo_2021} and LSND ~\cite{LSND:1996vlr} experiments have hinted at the existence of possible new physics beyond the standard model (BSM) for two decades. The discovery of one or more new particles would represent the first laboratory evidence of a new paradigm in high energy physics since the detection of neutrino mass through oscillations with profound implications for particle, astrophysics, and cosmology. As such, a great deal of effort has been focused on confirming, or rejecting, various interpretations of these anomalous excesses. These efforts include the construction of the MicroBooNE experiment \cite{MicroBooNE:2016pwy}, an 85 metric ton active volume liquid argon time projection chamber (LArTPC) situated in the Booster Neutrino Beam (BNB) at Fermilab~\cite{PhysRevD.79.072002}. The MicroBooNE experiment has been specifically designed to study the MiniBooNE anomaly as its primary goal. Both detectors were located in the same neutrino beam at a similar baseline.\\

 Possible origins of the MiniBooNE anomaly through more traditional hypotheses, such as underestimated backgrounds or systematic uncertainties~\cite{Brdar:2021ysi,Kelly:2022uaa}, have been unable to explain the anomaly and its resolution remains unclear. MicroBooNE has previously weighed in with multiple dedicated searches for both an anomalous excess of electron-like events~\cite{MicroBooNE:2021tya,MicroBooNE:2021pvo,MicroBooNE:2021nxr,MicroBooNE:2021wad,MicroBooNE:2024ryw}, a subsequent direct search for a light oscillating sterile neutrino in the BNB~\cite{MicroBooNE:2022sdp}, as well as a search for neutrino-induced neutral current (NC) production of the $\Delta$(1232) baryon resonance with subsequent $\Delta$ radiative decay~\cite{MicroBooNE:2021zai}, the dominant source of single photons predicted in both neutrino-argon and neutrino-carbon scattering below 1~GeV~\cite{Wang_2014}. While these placed strong bounds on their respective channels, MicroBooNE's first single-photon search was a model dependent search for NC $\Delta$ radiative decay specifically, and the strength of its result relied heavily on observing a proton in conjunction with the single-photon final state. Although this first analysis did contain a zero-proton sample, it had substantially more backgrounds. As such, any coherent-like single-photon final states without observable hadronic components were largely unconstrained. This possibility remains a viable and untested explanation of the MiniBooNE anomaly. \\

In recent years, there has been a rapidly growing interest in using neutrino experiments as a probe into dark sectors, hypothetical extensions of the standard model (SM) that introduce new particles and interactions below the electroweak scale. In many dark sector models, new unstable particles can be abundantly produced in neutrino-nucleus interactions through one or more portals, renormalizable terms added to the interaction Lagrangian that couple the dark sector weakly to the SM, such as the vector, scalar and neutrino portals \cite{Lanfranchi:2020crw}. It was highlighted that, alongside the long discussed electron and photon explanations of the MiniBooNE anomaly, if a dark sector particle decays to electron-positron ($e^+e^-$) pairs with $\mathcal{O}(100)$~MeV energies~\cite{Bertuzzo:2018itn,Ballett:2018ynz}, its signature can contribute to the excess of electron-like events observed by the MiniBooNE experiment. Although originally proposed to explain the MiniBooNE excess, the development of such models has since gained traction within the community and now forms a diverse class of theories that we are beginning to explore ~\cite{Ballett:2019pyw,Arguelles:2018mtc,Abdullahi:2020nyr,Abdallah:2020biq,Hammad:2021mpl,Datta:2020auq,Dutta:2020scq,Abdallah:2020vgg,Herwig:2023bnr,Acero:2022wqg}.  \\

\begin{figure}[h!]
\centering
    \includegraphics[width=0.47\textwidth]{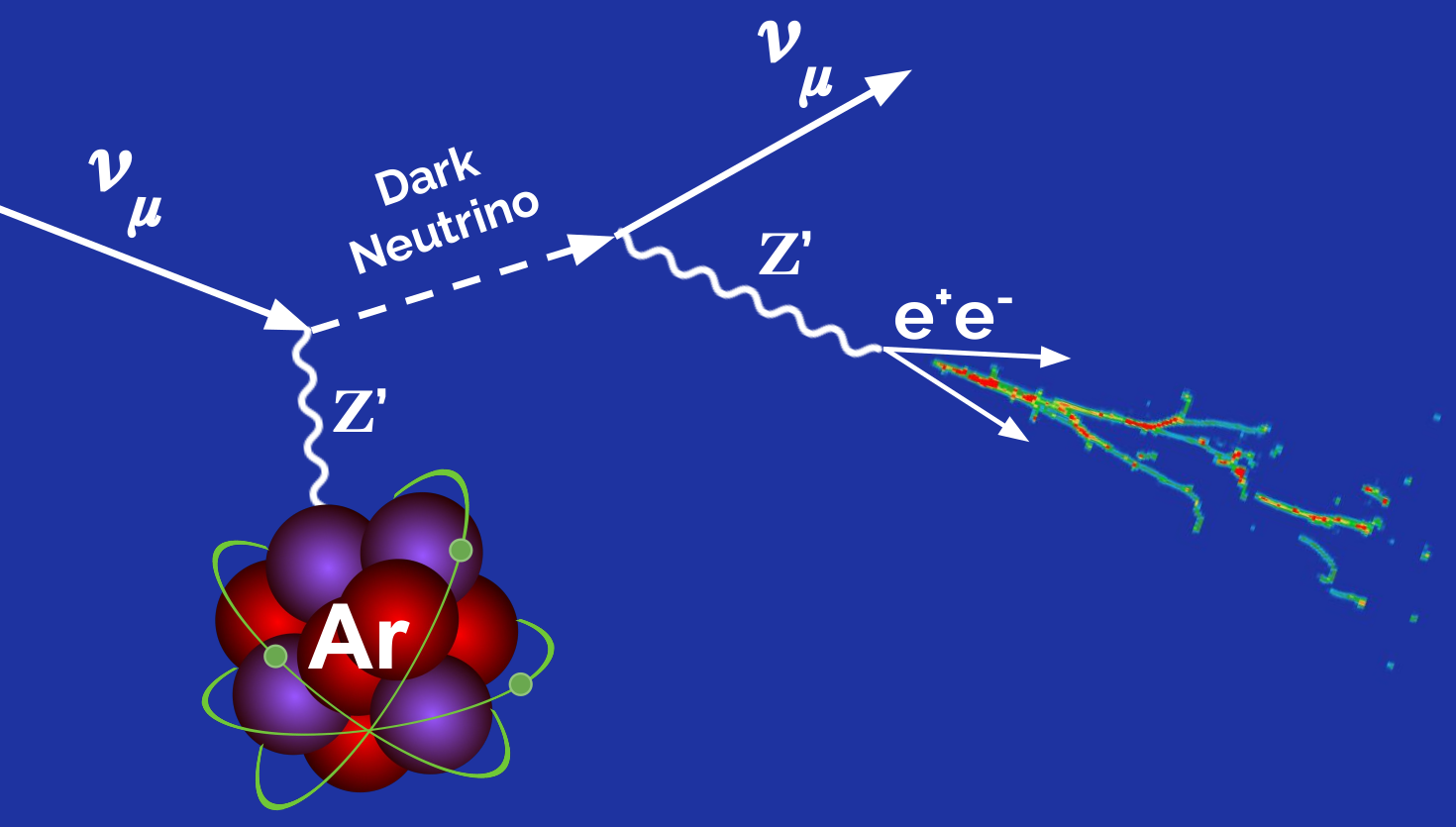}    
    \caption{An example of the dark sector models being probed. An incoming muon-neutrino from the BNB scatters off an argon target producing a dark neutrino. The subsequent decay of the dark neutrino leads to a visible $e^+e^-$ pair, mediated via a new dark gauge boson ($Z^\prime$). A simulated event is also shown behind the $e^+e^-$ arrows to represent what an $e^+e^-$ pair would actually look like in our LArTPC detector after forming electromagnetic showers.}
    \label{fig:cartoon1}
\end{figure}

This letter describes the first experimental search for dark sector $e^+e^-$ explanations of the MiniBooNE anomaly and complements and improves on previous phenomenological work~\cite{Arguelles:2018mtc,Brdar:2020tle,Arguelles:2022lzs}. We focus on the scenario in which active neutrinos from the BNB scatter to produce one or two dark neutrinos, which subsequently decay into missing energy and visible $e^+e^-$ pairs. It builds on prior MicroBooNE $e^+e^-$ searches utilizing the NuMI beamline~\cite{MicroBooNE:2023eef,MicroBooNE:2021sov,MicroBooNE:2023gmv,MicroBooNE:2025gpp}. While any dark sector may be quite rich phenomenologically, here a minimal $U(1)'$ gauge group is assumed with one or more dark neutrinos being the only particles carrying the group charge. The combination of neutrino mixing between active and dark neutrinos (the neutrino portal) as well as kinetic mixing between the dark gauge boson ($Z^\prime$) and the standard model photon (the vector portal) gives rise to the necessary phenomenology to explain the MiniBooNE anomaly. An example of how these dark sector models produce the visible $e^+e^-$ signal is shown in Fig.~\ref{fig:cartoon1} for a single dark neutrino. For dual dark neutrino models the outgoing muon neutrino would be replaced by a second, lighter, dark neutrino. To study these explanations, we use the phenomenological model as described in Ref.~\cite{Abdullahi:2023ejc} where we parameterize the interaction Lagrangian that couples the $Z^\prime$ boson to both neutral leptons and to standard model electromagnetic (EM) charge as:

\begin{equation}
    \mathcal{L}_\text{int}  \supset \sum^{3+n}_{i,j} g_D Z^\prime_\mu V_{ij} \overline{\nu_i}\gamma^\mu\nu_j - e \varepsilon  Z^\prime_\mu J^\mu_\text{EM},
\end{equation}
where $n$ is the number of dark neutrino states, $J^\mu_\text{EM}$ is the SM EM current with $\varepsilon$ as the kinetic mixing, $V_{ij} = \sum_{k=0}^n U^*_{ki}U_{kj}$ is the interaction vertex for dark current assuming dark neutrino states have a +1 charge under the $U(1)^\prime$ group. We adopt the same simplifying assumptions as Ref.~\cite{Abdullahi:2023ejc} in which dark neutrinos mix only with muon neutrinos. More details of the model and simplifying assumptions can be found in the Supplemental Materials. In this simplified regime  $|V_{\mu 4(5)}| = |U_{\mu 4(5)}|$, where latter being a more familiar parameter from the literature in heavy neutral leptons and the primary mixing element probed in this result.  \\

We consider models with one or two dark neutrinos. In the single dark neutrino model, the dark neutrino $\nu_4$ is long-lived unless it decays to an on-shell $Z^\prime$ boson via $\nu_4 \to \nu_\mu (Z^\prime \to e^+e^-)$. This need not be the case for $\nu_5$ in the dual dark neutrino models since it decays via $\nu_5 \to \nu_4 (Z^\prime \to e^+e^-)$. While the decay rate for $\nu_4$ is proportional to the small coupling between active and dark neutrinos, the decay rate for $\nu_5$ is proportional to an $\mathcal{O}(1)$ coupling between the dark sector $\nu_5$ and $\nu_4$. This allows us to explore regions of parameter space where the $Z^\prime$ is heavy ($m_{Z'} \gtrsim O(500)$~MeV) while still targeting dark neutrinos that have a large probability to decay inside the MicroBooNE detector after the initial neutrino interaction. \\ 

The kinematics of the decay process is also different between single and dual dark neutrino models. 
The latter depends on a key dimensionless parameter $\Delta = (m_5 - m_4)/m_4$, which controls the visible energy release in the $\nu_5\rightarrow \nu_4e^+e^-$ decays of interest. When $\Delta\gg1$, the decay kinematics is reduced to that of a single dark neutrino scenario, while for $\Delta<1$, the energy release into the final state $e^+e^-$ is suppressed. A further description of the class of models, as well as its ability to describe the MiniBooNE anomaly, can be found in the Supplemental Materials as well as references ~\cite{Abdullahi:2023ejc,Ballett:2019pyw,Abdullahi:2022cdw} and references therein with further discussions of UV completions of this phenomenological model in \cite{Ballett:2019pyw,Abdullahi:2020nyr,Bertuzzo:2018ftf}. \\

MicroBooNE uses a custom tune~\cite{PhysRevD.105.072001} of the \textsc{genie} neutrino event generator v3.0.6~\cite{Andreopoulos:2009rq,GENIE:2021zuu} to simulate backgrounds from neutrino-argon interactions. The BSM dark sector $e^+e^-$ signal events were  generated with the open-source \textsc{DarkNews}~\cite{Abdullahi:2022cdw} generator that is integrated into the first stage of MicroBooNE's LArTPC simulation chain. The \textsc{DarkNews} generator output was also validated with a GENIE implementation of the same single dark neutrino phenomenological model. Depending on the dark gauge boson mass, the initial scattering process on the nucleus can be predominantly coherent or incoherent, with the latter potentially resulting in multiple visible hadronic final states, in addition to the $e^+e^-$ pair. This analysis focuses solely on coherent scattering with no associated hadronic component for multiple reasons:  (a) this is the least constrained channel by MicroBooNE's prior results, (b)  the \textsc{DarkNews} generator lacks a detailed treatment of the complex nuclear response of the incoherent regime, and (c) when moving from carbon ($Z_\text{C}=6$, MiniBooNE's primary target) to argon ($Z_\text{Ar}=18$, MicroBooNE's target) we expect a substantial enhancement to coherent scattering that naively scales as atomic number squared, $(Z_\text{Ar}/Z_\text{C})^2 =9 $,  helping MicroBooNE's sensitivity to remain competitive by compensating for its lower active mass of 85 metric tonnes compared to MiniBooNE's 818 metric tonnes. \\

This coherent scattering tends to produce very forward-going dark neutrinos due to the low momentum exchange with the argon nucleus. When considering the subsequent dark neutrino decay, the model parameter space is broadly split into two regimes: for very prompt decays, with proper lifetime $c\tau << 10 $ cm,  the scattering on argon ($\approx 1.38 \text{~g/cm}^3$) and subsequent decay both  occur  predominately inside the detector, and when the proper lifetime grows,  $c\tau >> 10 $ cm, the portion of dark neutrino decays originating from neutrino scattering on argon inside the detector is dwarfed by the flux of long-lived dark neutrinos from neutrino scattering on the compacted soil and clay ($\approx 2.25 \text{g/cm}^3$ see~\cite{Baker:1983mj,dirt2,Cossairt:2010zz} and the Supplemental Materials) upstream of the detector and subsequently entering the MicroBooNE TPC. Both regimes are crucial for probing this class of models and are fully simulated in this analysis.  \\

A high statistics sample of dark sector signal events were generated and passed through MicroBooNE's simulation and reconstruction suite. This sample was distributed to cover the  heavy ($m_{Z^\prime} > m_{4(5)}$) and light ($m_{Z^\prime} << m_{4(5)}$) kinematic regions of interest of, as well as single and dual dark neutrino scenarios. Due to the high dimensionality of the dual dark neutrino parameter space, these simulated events were then reweighed to any point in our target model parameter space based on underlying true kinematic distributions. \\

The selection of $e^+e^-$ events and rejection of cosmic and neutrino backgrounds is built upon the framework developed for MicroBooNE's first generation NC $\Delta$ radiative search~\cite{MicroBooNE:2021zai}. Utilizing the Pandora pattern recognition reconstruction framework~\cite{ub_pandora}, ionization charge hits are first clustered and matched across three 2D projected views of the MicroBooNE active TPC volume into 3D reconstructed objects. These are then classified as tracks or showers based on a multivariate classifier score and aggregated into candidate neutrino interactions. While sufficiently overlapping $e^+e^-$ pairs are primarily reconstructed as single showers and are practically identical to true photons in our detector, $e^+e^-$ pairs with wider opening angles can be reconstructed as two-shower or one-shower plus one-track events. This analysis utilizes all three of these expected event topologies and is the first search for anomalous events consistent with MiniBooNE that targets events with two connected showers. \\

Once reconstructed, a series of four boosted decision trees (BDTs) are trained using the XGBoost~\cite{Chen:2016} algorithm to reject various background categories while selecting dark sector $e^+e^-$ events: 
\begin{itemize}
    \item Cosmic BDT: As a surface detector, there is a need to reject cosmogenic activity that enters in time with the TPC readout. It is trained on cosmic ray data events collected when no neutrino beam was present.
    \item Charged Current $\nu_\mu$ BDT: Although the most common interaction in the 99.5\% $\nu_\mu$ BNB, CC $\nu_\mu$ events are not a major background due the BDT being  able to easily identify and veto long muon tracks \cite{MicroBooNE:2021ddy}.  
    \item Charged Current $\nu_e$ BDT:  As the primary source of non-photon EM showers, it is vital to reject CC $\nu_e$ interactions to build confidence that selected events are truly $e^+e^-$. This BDT relies heavily on the calorimetric capabilities of a LArTPC \cite{MicroBooNE:2019rgx}, separating electrons from photons and $e^+e^-$ pairs by their energy deposition  $(dE/dx)$ at the start of the showering process.
    \item Neutral Current $\pi^0$ BDT: Neutral pions are by far the largest background to any photon or $e^+e^-$ search. If one daughter photon from a $\pi^0\rightarrow\gamma\gamma$ decay is not reconstructed or leaves the detector, the remaining shower can be indistinguishable from our signal.
\end{itemize}

\begin{figure}[h!]
    \centering
    \vspace{-0.5cm}
    \includegraphics[width=\linewidth]{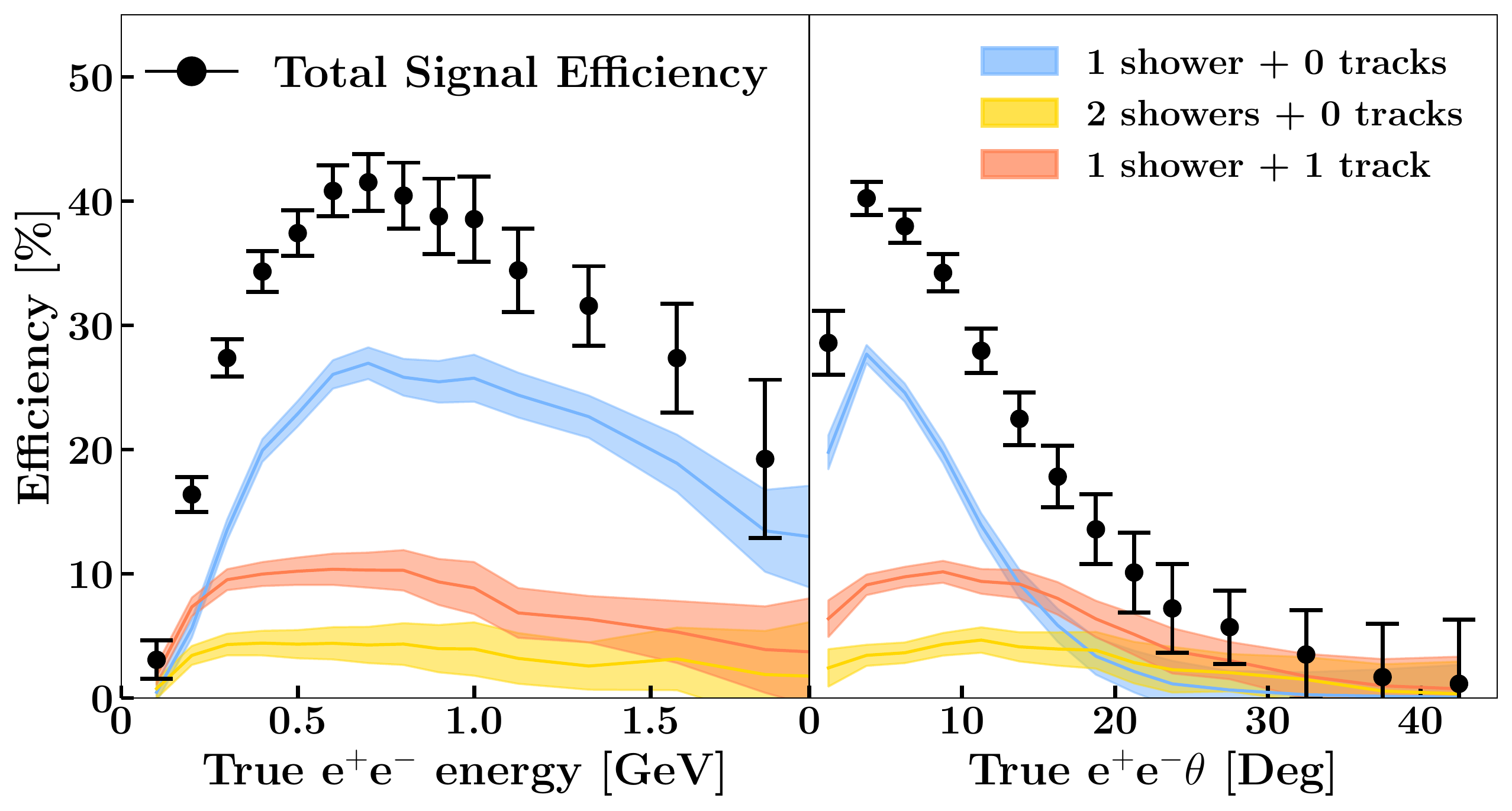}
    \caption{Relative signal efficiency as a function of (a) true $e^+e^-$ energy and (b) true $e^+e^-$ $\theta$ (angle with respect to neutrino beam) for a sample of dark sector signal events generated uniformly inside the TPC and covering a wide range of model parameter space.  Shown also is the how the efficiency varies when one splits the total efficiency into the three reconstructible topologies this analysis targets. Total efficiency is highly dependent on exact model parameters chosen which influence these variables as well as the $e^+e^-$ energy asymmetry. For more information see the Supplemental Materials.     }
    \label{fig:eff}
\end{figure}

As this analysis is focused on coherent BSM signals, we implement recently developed tools to search for low energy deposits of charge (down to 10 MeV) in the backwards direction of the reconstructed object to help veto any events with evidence of proton activity~\cite{abratenko2025searchneutralcurrentcoherent}. Events with BDT scores above a chosen threshold are kept, optimizing signal efficiency across the entire dark sector parameter space. The total efficiency for selecting dark sector $e^+e^-$ events varies across the model parameter space but is typically in the range $20-40\%$. Figure~\ref{fig:eff} shows a representative efficiency as a function of true $e^+e^-$ energy and overall angle with respect to the neutrino beam. \\

Prior to unblinding the signal region, the analysis underwent a series of validations in background rich sideband channels designed to ensure our modeling of NC $\pi^0$ backgrounds was sufficient. In all sidebands studied, we observe good agreement between data and our simulation prediction within assigned uncertainties. In addition to these dedicated background studies, a high statistics two-shower sideband in which both daughter photons of $\pi^0$ decays were successfully reconstructed is also employed to constrain the large cross-section uncertainties associated with NC $\pi^0$ production. For more details of this sideband and the constraint see the Supplemental Materials and~\cite{MicroBooNE:2021zai,MicroBooNE:2022zhr}. \\

\begin{table}[h!]
\centering
\begin{tabular}{|c|c|c|c|c|}
\cline{1-2}  \cline{4-5}  
\multicolumn{2}{|c|}{\textbf{Background Events}} &&\multicolumn{2}{c|}{\textbf{Uncertainty Breakdown}} \\ \cline{1-2}  \cline{4-5}  
{NC 1$\pi^0$} & 42.3 && {Cross-Section} &  24.2\% \\ \cline{1-2}  \cline{4-5}  
{Out of TPC} & 11.8 && {Detector} & 19.9\%   \\ \cline{1-2}  \cline{4-5}  
{CC 1$\pi^0$} & 6.9 && {Flux} & 8.0\%  \\ \cline{1-2}  \cline{4-5}  
{Cosmic} & 5.2 &&  {MC Stat.} & 5.5\% \\ \cline{1-2}  \cline{4-5}  
{Other}  & 10.2  && {Geant4} & 1.2\% \\ \cline{1-2} \cline{4-5}  \hline \hline 
\multicolumn{5}{|c|}{{Unconstrained Background:} 76.4 $\pm$ 25.0 (32.8\%) } \\ \hline
\multicolumn{5}{|c|}{\textbf{Constrained Background:} 69.7 $\pm$ 17.3 (24.9\%) } \\ \hline \hline
\end{tabular}
\caption{Breakdown of expected background events and the total associated systematic uncertainties for the final $e^+e^-$ selection both before and after applying the high statistics NC $\pi^0$ constraint. Percentages refer to the uncertainty on the predicted number of events.}
\label{tab:breakdown}
\end{table}

The effect of five categories of systematic uncertainties on our predictions are included: neutrino flux, neutrino interaction cross-sections, secondary interactions of hadrons outside of the target nucleus during \textsc{Geant4} simulation \cite{GEANT4:2002zbu}, the detector response model, and finite Monte-Carlo simulation statistics. The effect of the detector response is estimated in the same manner \cite{MicroBooNE:2021roa} as our prior NC $\Delta$ radiative analysis~\cite{MicroBooNE:2021zai}. The exact breakdown of expected backgrounds after selection as well as the effect of the NC $\pi^0$ constraint on the predicted background and its uncertainty can be found in Table.~\ref{tab:breakdown}. As with prior MicroBooNE photon searches, the vast majority of remaining backgrounds are NC $\pi^0$ events in which one photon either left the detector ($\approx 21\%$) or failed to be reconstructed ($\approx 79\%$), leaving a single shower to mimic the $e^+e^-$ signal. \\

\begin{figure}[h!]
    \centering
    \includegraphics[width=\linewidth]{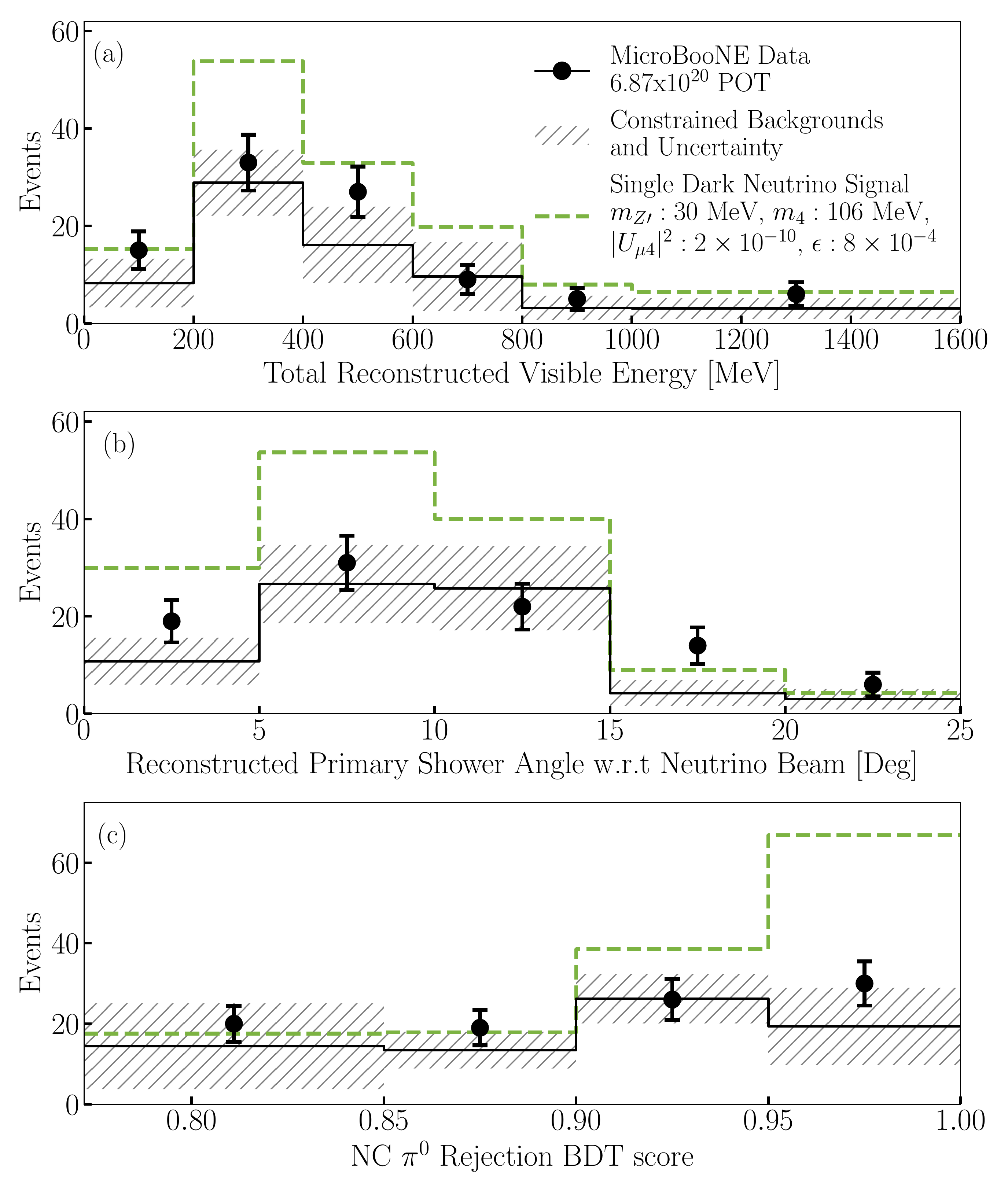}
    \caption{Final selected data and constrained background prediction for the $e^+e^-$ signal region. Shown is the reconstructed total energy (a), the angle the primary reconstructed shower makes with respect to the incoming neutrino beam (b) and the NC $\pi^0$ BDT rejection score (c). Shown also in red is a single dark neutrino example signal stacked on top of the background prediction, chosen to lie just within our predicted 95\% C.L. sensitivity, with $m_4=106$ MeV, $m_{Z^\prime}=30$ MeV, $|U_{\mu4}|^2=2.0\times10^{-10}$, and $\epsilon = 8 \times10^{-4}$. Equivalent spectra for pre-constrained backgrounds can be found in the Supplemental Materials.}
    \label{fig:data_mc}
\end{figure}

After unblinding data corresponding to an exposure of $6.87 \times 10^{20}$ protons-on-target (POT) we observe 95 $e^+e^-$ candidate events in the signal region, consistent with a constrained background-only prediction of 69.7 $\pm$ 17.3 at the $1.5\sigma$ level. For the three signal event topologies studied (single-shower, one-shower plus one-track, two-shower) we observe 52, 26, and 17 data events with constrained background predictions of $47.7\pm11.7$, $15.5\pm7.9$, and $9.5\pm4.6$ respectively. Distributions showing observed spectra in total reconstructed visible energy (defined as the sum of energies across all reconstructed showers and tracks), reconstructed primary shower angle, and NC $\pi^0$ background rejection BDT can be seen in Fig.~\ref{fig:data_mc}. In all cases we see agreement between observed data and our background-only prediction within assigned uncertainties. \\ 

\begin{figure}[h!]
    \centering
    \subfloat[Single dark neutrino scenario, $\varepsilon = 8\times10^{-4}$.]{
        \includegraphics[trim=0 0 0 0, clip, width=\linewidth]{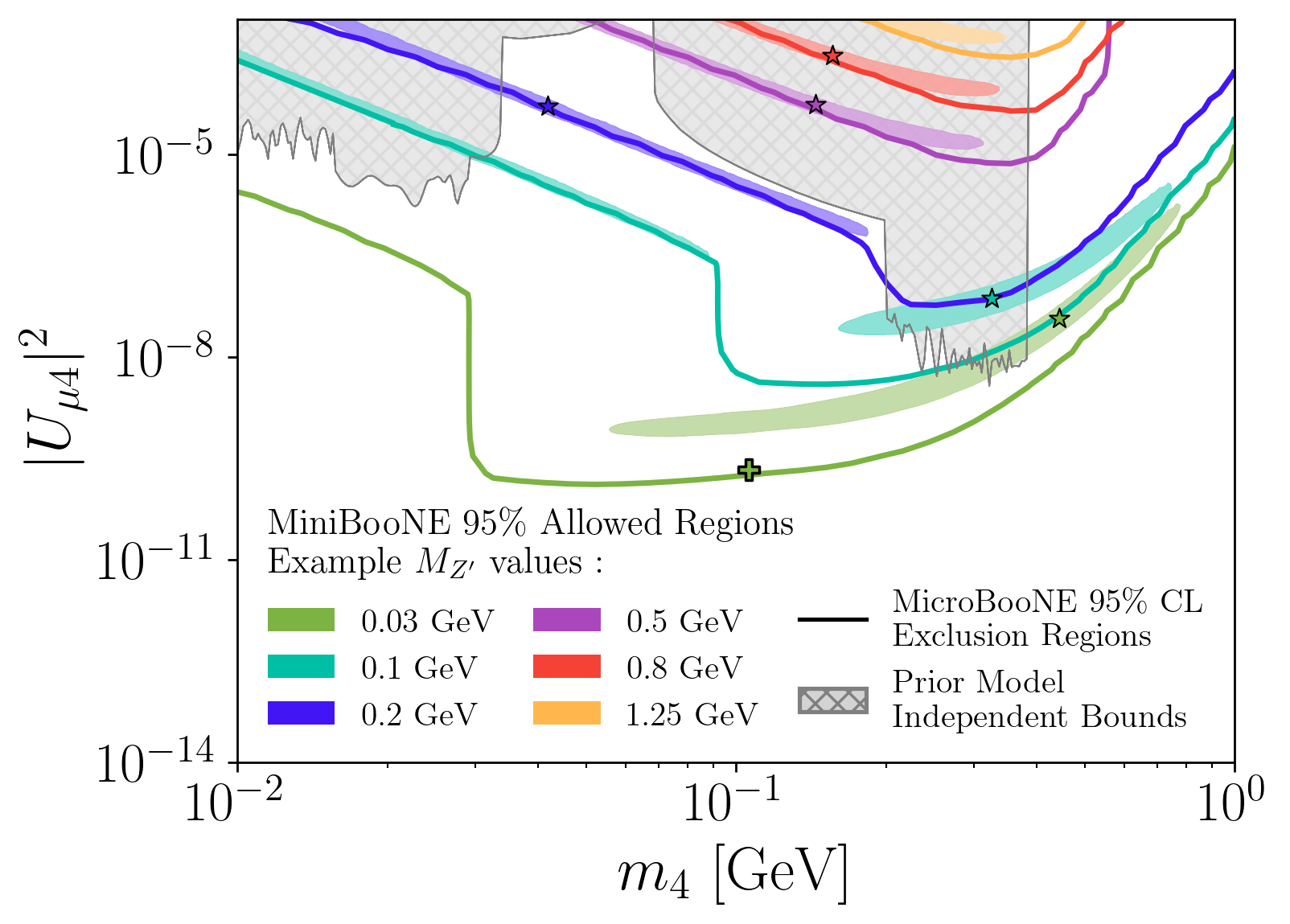}
    }
    
    \subfloat[Dual dark neutrino scenario, $\Delta = 1$, $\varepsilon = 8\times10^{-4}$.]{
        \includegraphics[trim=0 0 0 0, clip, width=\linewidth]{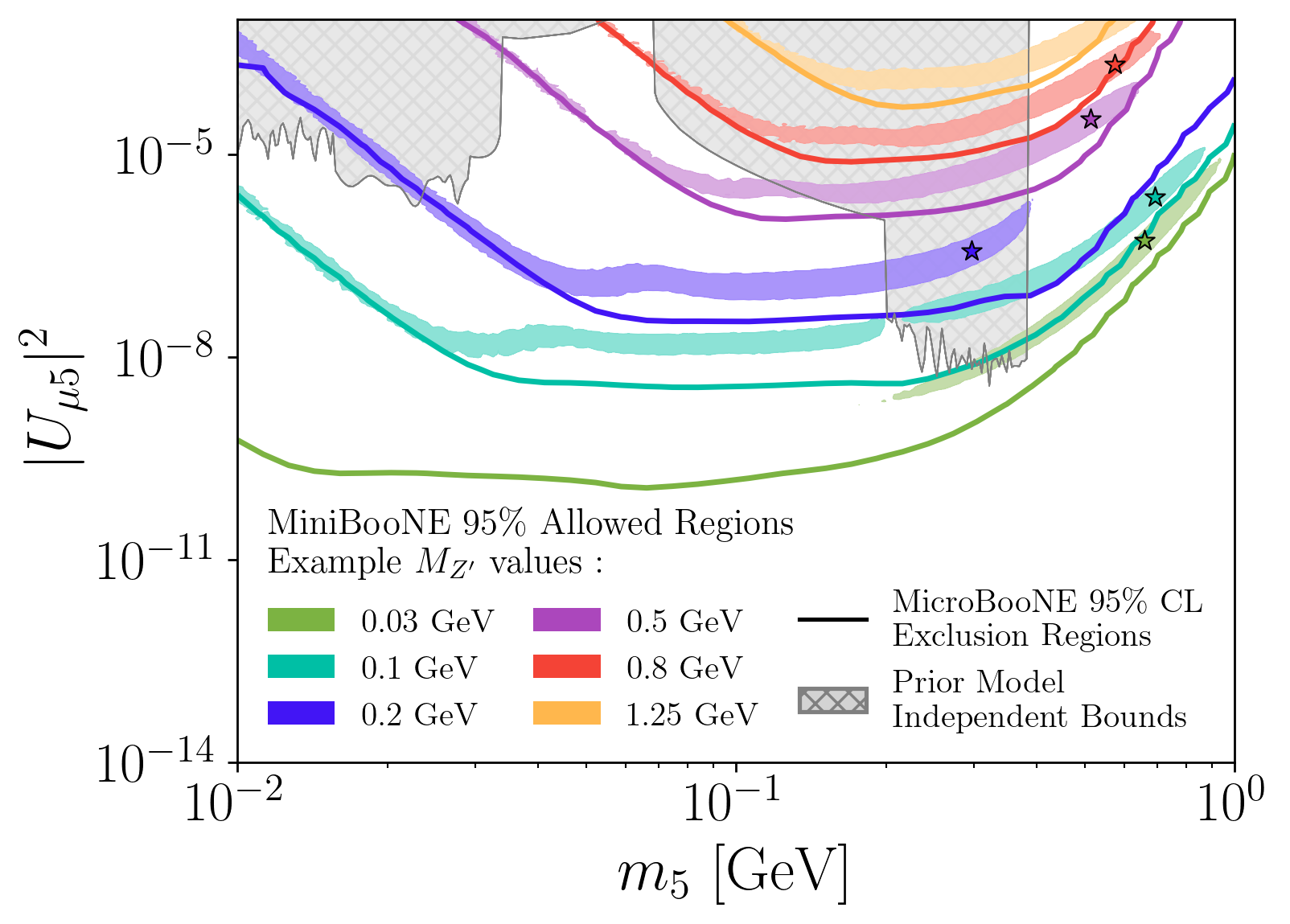}
    }
    \caption{The resulting MicroBooNE 95\% C.L. exclusion contours for a single dark neutrino model (a) and representative $\Delta=1$ dual dark neutrino scenario (b). Each solid line represents MicroBooNE limits for one of six values of the dark gauge boson mass ($m_{Z^\prime}$) varying from 30 MeV to 1.25 GeV, with everything above the solid line being excluded at above the 95\% C.L. The corresponding MiniBooNE allowed regions for each dark gauge boson mass indicate the 95\% C.L. preferred region of parameter space that offers a viable explanation to the MiniBooNE anomaly, taken from~\cite{Abdullahi:2023ejc}, with each star showing the MiniBooNE best fit for that single value of $M_{Z^\prime}$. The green plus in (a) indicates the example model point used in Fig.~\ref{fig:data_mc}. The gray solid regions indicate model-independent limits on dark neutrinos~\cite{Hayano:1982wu,PIENU:2019usb,Daum:1987bg,NA62:2021bji,BNL-E949:2009dza,Fernandez-Martinez:2023phj} prior to this result. The 30 MeV $Z^\prime$ contour in (a) corresponds to the dark neutrino model introduced in Ref.~\cite{Bertuzzo:2018itn}.}
    \label{fig:results}
\end{figure}

 Studies of the sensitivity to these dark sector $e^+e^-$ signals are performed using the total event rate, 
 total reconstructed visible energy, reconstructed angle, and NC $\pi^0$ rejection BDT score. The most sensitive variable tested, the NC $\pi^0$ rejection BDT score, is presented here in more detail. The selected $e^+e^-$ candidates were fit alongside the two-shower NC $\pi^0$ sidebands in order to constrain backgrounds using the combined Neyman-Person (CNP) $\chi^2$ test statistic~\cite{Ji:2019yca}. This test statistic includes all systematic uncertainties discussed above through a covariance matrix. Excellent agreement with the constrained background-only prediction was observed with a $\chi^2_\text{CNP}$ of 3.08 for 4 degrees-of-freedom (\emph{d.o.f.}). Given this agreement we calculate limits from a fit to the NC $\pi^0$ rejection BDT score on both single and dual dark neutrino models of dark sector production in Fig.~\ref{fig:results}. We present the limits for six representative values of the mass of the dark gauge boson, $Z^\prime$, and the dual dark neutrino result for fixed relative mass gap $\Delta=1$. This is both for simplicity of presentation and to allow direct comparison to the published MiniBooNE preferred regions of model parameter space that can explain the MiniBooNE low-energy anomaly~\cite{Abdullahi:2023ejc}. Limits are obtained for each $Z^\prime$ mass individually by placing cuts on $\Delta \chi^2_\text{CNP} < 5.99$, corresponding to a 95\% C.L assuming Wilk's theorem for 2 \emph{d.o.f}. For the majority of parameter space probed, MicroBooNE places world leading exclusion limits on this class of models, with the MicroBooNE 95\% confidence level exclusion limits ruling out this interpretation of the MiniBooNE anomaly. \\

This letter has presented the first dedicated analysis exploring $e^+e^-$ explanations for the short-baseline MiniBooNE anomaly, as suggested by non-minimal dark sector models. 
By expanding beyond past single-photon searches at MicroBooNE to a broader $e^+e^-$ signal topology containing also two-shower and one-shower, one-track events, a significant increase in efficiency is achieved for very forward-going, coherently-produced $e^+e^-$ pairs. 
For these models predicting such a signature, we find no evidence of a signal consistent with MiniBooNE's observation. This result holds for both scenarios involving light $Z^\prime$ mediators and those with heavier $Z^\prime$ states. Our analysis significantly constrains the majority of the model phase space motivated by the MiniBooNE anomaly. Nevertheless, alternative dark sector models with different mediators, such as scalar mediators or those dominated by incoherent scattering leading to distinct energy and angular distributions, are not constrained by this result and remain promising avenues for future exploration.

\vspace{2cm}
\begin{acknowledgments}  
This document was prepared by the MicroBooNE collaboration using the
resources of the Fermi National Accelerator Laboratory (Fermilab), a
U.S. Department of Energy, Office of Science, Office of High Energy Physics HEP User Facility.
Fermilab is managed by Fermi Forward Discovery Group, LLC, acting
under Contract No. 89243024CSC000002.  MicroBooNE is supported by the
following: 
the U.S. Department of Energy, Office of Science, Offices of High Energy Physics and Nuclear Physics; 
the U.S. National Science Foundation; 
the Swiss National Science Foundation; 
the Science and Technology Facilities Council (STFC), part of the United Kingdom Research and Innovation; 
the Royal Society (United Kingdom); 
the UK Research and Innovation (UKRI) Future Leaders Fellowship; 
and the NSF AI Institute for Artificial Intelligence and Fundamental Interactions. 
Additional support for the laser calibration system and cosmic ray tagger was provided by the Albert Einstein Center for Fundamental Physics, Bern, Switzerland. We also acknowledge the contributions of technical and scientific staff to the design, construction, and operation of the MicroBooNE detector 
as well as the contributions of past collaborators to the development 
of MicroBooNE analyses, without whom this work would not have been 
possible.  For the purpose of open access, the authors have applied 
a Creative Commons Attribution (CC BY) public copyright license to 
any Author Accepted Manuscript version arising from this submission.

\end{acknowledgments}

\bibliography{apssamp}

\end{document}